\documentclass[a4paper,english,12pt]{article}

\usepackage{amssymb}
\usepackage{amsmath}
\usepackage{epsfig}
\usepackage{overcite}
\makeatletter \renewcommand\@biblabel[1]{$^{#1}$} \makeatother

\newcommand{\R}{\mathbb{R}}
\newcommand{\N}{\mathbb{N}}

\begin{document}

\title{Bounds on the {\em dragging rate} and on the {\em rotational mass-energy} 
in slowly and differentially rotating relativistic stars}
\author{M. J. Pareja\\
Dept. de F\'{\i}sica Te\'orica II, Facultad de Ciencias F\'{\i}sicas,\\
Universidad Complutense de Madrid, E--28040 Madrid, Spain\\
{\em e-mail: mjpareja@fis.ucm.es}
}
\date{}

\maketitle
\begin{abstract}
For relativistic stars rotating slowly and differentially
with a positive angular velocity,
some properties in relation to the positiveness of the
rate of rotational dragging and of the angular momentum density
are derived. Also, a new proof for the bounds on the rotational
mass-energy is given.
\end{abstract}
PACS numbers: 04.40.Dg, 97.10.Kc, 02.30.Jr

\section*{\normalsize I. INTRODUCTION}

In the prescription for calculating a slowly and differentially rotating 
relativistic stellar configuration the field equations are expanded in powers of
a {\em fluid angular velocity parameter} and the perturbations (around a non-rotating
configuration) are calculated by retaining only first and second order terms. 
Hartle\cite{H1} has derived these equations of structure in the rigidly rotating case.
However, at first order the only effect of the rotation is to drag the inertial frames; 
at second order it also deforms the star. In fact, the first order equations
of structure reduce to the time-angle field equation component (to first order), which is
a partial differential equation, linear in the {\em angular velocity of cumulative dragging} 
(or {\em dragging rate}). 

This linearity in the dragging rate potential persuades us to write 
that equation in appropriate coordinates
---in order to avoid the coordinate singularity occurring on the axis of 
rotation in spherical polar coordinates, generally used in the slow rotation approximation--- 
so that in the new coordinates the equation writes in a ``regular'' form 
as an elliptic equation with bounded coefficients, and to apply a minimum principle
for generalized supersolutions in the whole domain (interior and exterior of the star).
Making use of the asymptotic flatness condition, this will lead us
directly to the positivity of the dragging rate, provided that the distribution of 
angular velocity of the fluid is non-negative everywhere (and non-trivial) and that 
we start from a reasonable unperturbed (non-rotating) stellar model satisfying the
weak energy condition.
This and the positiveness of other quantities also linear in the 
angular velocity, like the {\em angular momentum density},
will be the purpose of this paper.

\vspace*{1mm}

The {\em rotational mass-energy}, derived by Hartle\cite{H4}, 
although accurate to second order in the angular velocity, 
involves only quantities which can be calculated from the
first order structure equation (time-angle component of the Einstein equations) 
as well. A proof of the positivity and an upper bound on this rotational energy is
given in the same paper \cite{H4}, however using an expansion in eigenfunctions.
We shall give here a much simpler proof of these bounds, without using that
expansion, and, hence, avoiding the non-trivial mathematical problems on the
existence of these eigenfunctions.

\vspace*{1mm}

The paper is organized as follows. After a description of the relativistic 
rotating stellar model in Sec.~II, and a brief revision of the concepts of
angular momentum density and rate of rotational dragging in Sec.~III, 
in Sec.~IV we concentrate on the slow rotation approximation, particularly
on the first order perturbations of the metric (linear correction of the
dragging rate, with description of the unperturbed (zero order) configuration), 
and explicit expressions for the expansions of the angular momentum density
and of the rotational mass-energy are derived. In the same section the null
contribution (at first order in the angular velocity) of the integrability 
condition of the Euler equation is discussed, and the time-angle 
component of the Einstein equations (to first order) 
is written in appropriate coordinates, as a background allowing to
apply a minimum principle and obtain the first of the properties mentioned 
above and proved in Sec.~V, and consequences of that one. 
Apart from this, as an independent result, an alternative
proof of the bounds on the rotational energy is given.
Finally, in Sec.~VI, concluding remarks are briefly stated.

\section*{\normalsize II. THE RELATIVISTIC ROTATING STELLAR MODEL}

The spacetime of a rotating relativistic star is represented by a
Lorentzian $4$-manifold $(\cal{M},{\mathbf g})$ which satisfies the following

\subsection*{A. Assumptions}

\begin{enumerate}

\item[i.] the spacetime is stationary in time and axially symmetric,
which means that ${\mathbf g}$ admits two global Killing vector fields,
a time-like future-directed one, ${\boldsymbol \xi}$, and a space-like one, with
closed trajectories, ${\boldsymbol \eta},$ except on a time-like 2-surface
(defining the axis of rotation) where ${\boldsymbol \eta}$ vanishes;

\item[ii.] the spacetime is asymptotically flat; in particular,
${\mathbf g}({\boldsymbol \xi},{\boldsymbol \xi}) \to -1, \
{\mathbf g}({\boldsymbol \eta},{\boldsymbol \eta}) \to +\infty,$ and ${\mathbf g} 
({\boldsymbol \xi},{\boldsymbol \eta}) \to 0$
at spatial infinity (the signature of the metric ${\mathbf g}$ being $(- + + +));$

\item[iii.] the matter ---confined in a compact region in the space
(interior), with vacuum on the outside, so that (ii) holds--- is
perfect fluid, and therefore the energy-momentum tensor (source of the
Einstein equations) is written as
$$
{\mathbf T} = (\varepsilon + p) {\mathbf u}^{\flat} \otimes {\mathbf u}^{\flat} +
p\, {\mathbf g} \, ,
$$
where $\varepsilon$ and $p$ denote the energy density and the pressure
of the fluid, respectively; and ${\mathbf u^{\flat}}$ denotes the 1-form equivalent
to the $4$-velocity of the fluid ${\mathbf u}$ \ (in the exterior ${\mathbf T} \equiv 0;$
hence, $\varepsilon + p = p = 0$ there);

\item[iv.] the fluid velocity is azimuthal (non-convective)
({\it circularity condition}), i.e.
$$
{\mathbf u}^{\flat} \wedge {\boldsymbol \xi}^{\flat} \wedge {\boldsymbol \eta}^{\flat} = 0 \, ;
$$

\item[v.] $(\cal{M},{\mathbf g})$ satisfies Einstein's field equations ${\mathbf G} =
8\pi{\mathbf T}$ for the energy-momentum tensor ${\mathbf T}$ of a perfect fluid (iii),
where ${\mathbf G} \equiv {\mathbf {\mbox Ric}} - 1/2 \, R \, {\mathbf g}$ denotes the 
Einstein tensor ---equations which can also be written in the form 

\begin{equation}
{\mathbf{\mbox Ric}} = 
8\pi ({\mathbf T} - \frac{1}{2} {\mbox tr}({\mathbf T}) \, {\mathbf g}) \, ;
\label{Einstein_eqs}
\end{equation}

\item[vi.] $\varepsilon$ and $p$ satisfy a barotropic (one-parameter)
equation of state, $\varepsilon = \varepsilon(p);$

\item[vii.] $\varepsilon + p \ge 0$ ({\em weak energy condition} for perfect fluid,
assuming $\varepsilon \ge 0)$ \cite{Wald};


\item[viii.] the metric functions are essentially bounded.

\end{enumerate}

\subsection*{B. Form of the metric}

Assumptions (i) and (ii) imply that the two Killing fields commute, $[{\boldsymbol \xi},
{\boldsymbol \eta}] = 0$ \cite{C}, which is equivalent to the existence of coordinates
$x^0 \equiv t$
and $x^1 \equiv \phi$ such that ${\boldsymbol \xi} \equiv \partial_t$ and
${\boldsymbol \eta} \equiv \partial_{\phi};$ moreover, by the circularity
condition (iv), the spacetime ${\mathbf g}$ admits $2$-surfaces orthogonal to
the group orbits of the Killing fields (orthogonal transitivity) \cite{K-T}.
We may then choose the two remaining coordinates $(x^2,x^3)$ in one of these
$2$-surfaces and carry them to the whole spacetime along the integral curves
of ${\boldsymbol \xi}$ and ${\boldsymbol \eta};$ accordingly, the metric can be written
in the form

\begin{eqnarray}
ds^2 & \equiv & g_{\alpha \beta} dx^{\alpha} dx^{\beta}= \nonumber\\
 & = & g_{tt} dt^2 + 2g_{t \phi} dt d\phi + g_{\phi \phi} d\phi^2 +
g_{22}(dx^2)^2 + 2 g_{23}dx^2 dx^3 + g_{33}(dx^3)^2,\nonumber
\end{eqnarray}

\noindent where the metric coefficients are independent of
the time $x^0 \equiv t$ and azimuthal $x^1 \equiv \phi$
coordinates corresponding to the two Killing fields; that is,
$g_{\alpha \beta} = g_{\alpha \beta}(x^2,x^3).$

When solving Einstein's field equations it is convenient to specify
the coordinates $x^2$ and $x^3$ in such a way as to simplify the task;
a particular choice, usually made when studying slowly rotating
configurations \cite{H1}, is the one which makes $g_{23} = 0$ and $g_{33} =
g_{\phi\phi} \sin^{-2} \! x^3.$
Hence $x^2$ and $x^3$ are chosen so that at large spatial distances
the asymptotically flat metric is expressed in terms of spherical polar coordinates
in the usual way.
In the resulting coordinate system, with the notation $x^2 \equiv r,
\ x^3 \equiv \theta$, and with new symbols for the metric functions,
the line element reads

\begin{equation}
ds^2 = -H^2 dt^2 + Q^2 dr^2 + r^2 K^2 [d\theta^2 + \sin^2 \!\theta
\, (d\phi - A dt)^2] \, ,
\label{metric}
\end{equation}

\noindent where $H, Q , K$ and $A$ are functions of $r$ and $\theta$ alone.
In these coordinates $(r \ge 0, \ \ 0 \le \theta \le \pi)$
the spatial infinity is given by $r \to \infty,$ and the axis of rotation
$(\partial_{\phi} \equiv {\boldsymbol \eta} = 0)$ is described by 
$\theta \to 0$ or $\pi$ \ $(r \ge 0).$ 


Notice that the function $A$ appears in the metric (\ref{metric})
as the non-vanishing of the $(t \phi)$ metric component of a rotating
configuration. $A$ is actually the {\it dragging rate} potential
(cf.~Sec.~III).

\subsection*{C. Differential rotation}

According to assumption (iv) of Subsec.~II.A ---circularity condition---,
the fluid $4$-velocity ${\mathbf u}$ has the form
$$
{\mathbf u} = u^t \partial_t + u^{\phi} \partial_{\phi} = u^t (\partial_t +
\Omega \partial_{\phi}), \ \ \ \ \mbox{where} \ \ \Omega \equiv \frac{u^{\phi}}{u^t} =
\frac{d\phi}{dt}
$$
is the angular velocity of the fluid measured in units of $t$,
i.e. as seen by an inertial observer at infinity whose proper time is the same
as the coordinate $t$ (observer $\partial_t),$ and $u^t$ is the
normalization factor, such that \ ${\mathbf g}({\mathbf u},{\mathbf u}) = -1,$ \ i.e. \ 
$(u^t)^{-2} = -(g_{tt} + 2 \Omega g_{t \phi} + \Omega^2 g_{\phi \phi}).$ 
The star's matter rotates then in the azimuthal direction $\phi.$

We consider a star rotating differentially, with a prescribed distribution of
angular velocity
$$
\Omega \equiv \Omega(x^2, x^3) \equiv \Omega(r,\theta) \, ,
$$
an essentially bounded function.
However, with the assumptions made (Subsec.~II.A), the rotation profile of the fluid
cannot be freely chosen, this shows up in the following. We consider the equation
of hydrostatic equilibrium, $T^{\alpha \beta}_{\ \ \ ;\beta} = 0$ \ (integrability 
conditions of the field equations), particularly, its part orthogonal to 
the fluid $4$-velocity ${\mathbf u}$, i.e. the Euler equation,

\begin{equation}
dp = - (\varepsilon + p)\, {\mathbf a} \, ,
\label{Euler_eq}
\end{equation}

\noindent where $\mathbf{a}$ is the $4$-acceleration of the fluid,
$\mathbf{a}^{\sharp} = \nabla_{{\mathbf u}} {\mathbf u},$ specifically, 

\begin{equation}
{\mathbf a} =
- d \ln u^t + u^t u_{\phi} \, d \Omega \, .
\label{acc}
\end{equation}

\noindent And the integrability condition of Eq.~(\ref{Euler_eq}), taking into account (vi) of 
Subsec.~II.A, $\varepsilon = \varepsilon(p),$ is \ $d \mathbf{a} = 0;$ following, from (\ref{acc}),
$d (u^t u_{\phi}) \wedge d \Omega = 0;$
in other words, the fluid angular velocity, $\Omega,$ is functionally related to the specific 
angular momentum times $u^t$,

\begin{equation}
u^t u_{\phi} = {\cal F} (\Omega) \, .
\label{Euler_intc}
\end{equation}

Nevertheless, as will be seen in Subsec.~IV.B, in the slow rotation approximation,
at first order in the angular velocity, Eq.~(\ref{Euler_intc}) is no
restriction on $\Omega(r,\theta).$

\section*{\normalsize III. ANGULAR MOMENTUM DENSITY AND DRAGGING RATE}

The total angular momentum of a rotating relativistic star can be defined \cite{H-Sh}
from the variational principle for general relativity ---for an isolated system which
is not radiating gravitational waves---, but this is shown \cite{H-Sh} to coincide
with the {\it geometrical} definition ---from the asymptotic form of the metric at
large space-like distances from the rotating fluid--- (analog to the ADM mass), which
for stationary and axisymmetric (asymptotically flat) spacetimes is given by the
Komar integral for the angular momentum \cite{K}, a surface integral, which,
when reformulated using the Gauss theorem and the Einstein equations, converts into
the volume integral over the interior

\begin{eqnarray}
J & = & \int_{\cal I} T_{\alpha}^{\ \beta} \eta^{\alpha} n_{\beta} dv \label{J} \\
& = & \int_{\cal I} T_{\phi}^{\ t} n_t dv 
= \int_{\cal I} T_{\phi}^{\ t} (- g)^{1/2} d^3 x \, , \label{Jh} 
\end{eqnarray}

\noindent where ${\mathbf T}$ is the energy-momentum tensor of perfect fluid,
${\boldsymbol \eta}$ is the Killing field corresponding to the axial symmetry,
${\mathbf n}^{\sharp}$ is the unit time-like and future pointing normal to the hypersurface
of constant $t,$ i.e. ${\mathbf n} = n_t \, dt,$ with $n_t > 0,$
and $dv$ is the proper volume element of the surface $t=$ const., i.e.
$\int_{\cal I} dv =$ Vol, the volume of the body of the star, ${\cal I} \equiv$ interior
of the star $(t=$ const.). ${\rm g} \equiv \det({\mathbf g}).$ 
The invariantly defined integrand of this volume integral (\ref{J}),
$T_{\alpha}^{\ \beta} \eta^{\alpha} n_{\beta},$ is what
one would naturally define  as {\it angular momentum density} ---coinciding
with the standard form in special relativity---, and can be calculated

\begin{eqnarray}
T_{\alpha}^{\ \beta} \eta^{\alpha} \, n_{\beta} & = & n_t \, T_{\phi}^{\ t} 
\label{amd} \\
& = & n_t \, (\varepsilon + p) u^t u_{\phi} \
\ \ \ [u_{\phi}= u^t (g_{t\phi} + \Omega g_{\phi \phi})] \nonumber\\
& = & n_t \, (\varepsilon + p) (u^t)^2 (g_{t \phi} + \Omega
g_{\phi \phi})
\nonumber \\
& = & n_t \, (\varepsilon + p) (u^t)^2 g_{\phi \phi} (\Omega - A), \ \ \ \
\mbox{with} \ \ n_t = H = \left( \frac{-g_{tt} g_{\phi \phi}
+ g_{t \phi}^2}{g_{\phi \phi}} \right)^{1/2} \, ,\nonumber
\end{eqnarray}

\noindent where $A$ is the metric function (cf.~(\ref{metric})) such that

\begin{equation}
g_{t \phi} = - A \ g_{\phi \phi} \, .
\label{A}
\end{equation}

It is remarkable that, since $n_t > 0, \ g_{\phi \phi} \ge 0,$ and we are
assuming the energy condition 
$\varepsilon + p \ge 0$ ((vii) in Subsec.~II.A), the sign of
the angular momentum density (\ref{amd}) is determined by the sign of the
difference $\Omega - A.$ 

The metric function $A$ is indeed the angular velocity of a particle which
is dragged along in the gravitational field of the star, as seen from a
non-rotating observer at spatial infinity $(\partial_t)$, so that it has
zero angular momentum relative to the axis, $p_{\phi} = 0,$
$$
\frac{d\phi}{dt} = \frac{p^{\phi}}{p^t} = \frac{g^{t \phi} p_t}{g^{tt} p_t}
= \frac{g^{t \phi}}{g^{tt}} = \frac{- g_{t \phi}}{g_{\phi \phi}}; \ \ \ \ 
g_{t \phi} + \left( \frac{d\phi}{dt} \right) g_{\phi \phi} = 0 ; \ \ \ \
\frac{d\phi}{dt} = A \, .
$$
\noindent $A$ is called {\it angular velocity of cumulative dragging}
(shortly called {\it dragging rate}) \cite{Bardeen,Thorne}. One of the purposes 
of this work is precisely to find appropriate bounds on the uniformly non-negative 
distribution of angular velocity, $\Omega \equiv \Omega(x^2,x^3) \ge 0,$ 
of a slowly differentially rotating star, so that $\Omega - A \ge 0$ holds;
from where the positivity of the angular momentum density (\ref{amd}) follows
(Property (c) in Sec.~V).

Observe, in the special relativistic limit 
$(g_{t \phi} \to 0,$ using coordinates $(x^2,x^3)$ which go at spatial
infinity to the usual flat coordinates, cf.~Subsec.~II.B), if the fluid rotates
uniformly with angular velocity $\Omega$ positive (negative), then the
angular momentum density, Eq.~(\ref{amd}), is uniformly positive (negative).
\section*{\normalsize IV. SLOWLY DIFFERENTIALLY ROTATING STARS. 
FIRST ORDER PERTURBATIONS}

By slow rotation we mean that the absolute value of the angular velocity is
much smaller than the critical value $\Omega_{crit} \equiv (M / R^3)^{1/2}$
(taking units $c = G = 1),$
where $M$ is the total mass of the unperturbed (non-rotating) configuration,
and $R,$ its radius; $|\Omega(x^2,x^3)| / \Omega_{crit} \ll 1.$
Thus, stars
which rotate slowly can be studied by expanding the Einstein field equations
for a fully relativistic differentially rotating star in powers of the
dimensionless ratio 
\begin{equation}
\frac{|\Omega_{max}|}{\Omega_{crit}} =: \mu \, ,
\label{mu}
\end{equation}  
\noindent where $|\Omega_{max}|$ is the maximum value of $|\Omega(x^2,x^3)|$ 
(at the interior of the star).

\subsection*{A. The metric and the energy density and pressure of the fluid}

We assume that the star is slowly rotating, with angular velocity
$$
\Omega(r,\theta) \equiv \Omega(x^2,x^3) = O(\mu) \, ,
$$
\noindent parameter $\mu$ given e.g. by (\ref{mu}).
Because the star (stationary in time and axially symmetric)
rotates in the $\phi$ direction ((iv) of Subsec.~II.A), a time reversal
$(t \to -t)$ would change the sense of rotation, as well as an inversion
in the $\phi$ direction $(\phi \to - \phi)$ would do. As a result, the
metric coefficients $H, Q$ and $K$ (in (\ref{metric})) and the energy
density will not change sign under {\it one of} \/ these inversions,
whereas $A$ will do. Therefore, an expansion of $H, Q$ and $K$,
as well as of the energy density, $\varepsilon,$ and, hence, of the pressure, $p,$
in powers of the angular velocity parameter $\mu$ will contain only
even powers, while an expansion of $A$ will have only odd ones.

Accordingly, at first order in the angular velocity, $O(\mu),$ 
the metric coefficients, and fluid energy density and pressure, are

\begin{eqnarray}
H & = & H_0 + O(\mu^2)\nonumber \\
Q & = & Q_0 + O(\mu^2)\nonumber \\
K & = & K_0 + O(\mu^2)\nonumber \\
& & \nonumber \\
\varepsilon & = & \varepsilon_0 + O(\mu^2)\nonumber \\
p & = & p_0 + O(\mu^2)\nonumber \\
& & \nonumber \\
\mbox{but}\ \ A & = & \omega + O(\mu^3), \label{omega}
\end{eqnarray}

\noindent where $H_0, Q_0$ and $K_0$ are the coefficients of the unperturbed
(non-rotating) configuration, and $\omega$ denotes the linear (first order) 
correction in $\mu$ of the dragging rate $A,$ so that, from Eq.~(\ref{A}),

\begin{equation}
g_{t \phi} = - \omega \ (g_{\phi \phi})_0 + O(\mu^3) \, .
\label{omegaexp}
\end{equation}

\noindent We shall keep here only the effects linear in the angular velocity.
The only first order $O(\mu)$ perturbation brought about by the rotation is the 
dragging of the inertial frames; the star is still spherical, because the
``potential functions'' which deform the shape of the star are $O(\mu^2).$

\subsubsection*{1. The (unperturbed) non-rotating configuration}

The starting non-rotating equilibrium configuration is described by
the spherically symmetric metric in the Schwarzschild form

\begin{eqnarray}
ds^2 & = & - e^{\nu(r)} \, dt^2 + e^{\lambda(r)} \, dr^2 + r^2(d\theta^2
+ \sin^2 \!\theta \, d\phi^2)\label{nrmetric}\\
& \equiv & - H_0^{\ 2} \, dt^2 + Q_0^{\ 2} \, dr^2 + r^2 K_0^{\ 2} \, (d\theta^2
+ \sin^2 \!\theta \, d\phi^2) \ \ (A_0 = 0)\nonumber,
\end{eqnarray}

\noindent with $\lambda(r),$
or equivalently, the mass $m(r)$
interior to a given radial coordinate $r,$ given by

\begin{equation}
1 - \frac{2m(r)}{r} = e^{-\lambda(r)} \, ,
\label{nr0}
\end{equation}

\noindent and $\nu(r),$ together with the pressure $p_0(r)$, and the energy density
$\varepsilon_0(r),$ solutions of the system of equations of general
relativistic hydrostatics, which for a non-rotating configuration are: 
the equation of hydrostatic equilibrium (Tolman-Oppenheimer-Volkoff equation),
\begin{equation}
\frac{d p_0}{dr}(r) =  - \, \frac{[\varepsilon_0(r) + p_0(r)]
[m(r) + 4 \pi r^3 p_0(r)]}{r^2[1- 2m(r) / r]} \, ,
\label{nr1}
\end{equation}
the mass equation, 
\begin{equation}
\frac{d m}{dr}(r) = 4 \pi r^2 \varepsilon_0(r) \, ,
\label{nr3}
\end{equation}
and the source equation for $\nu,$
\begin{equation}
\frac{d \nu}{dr}(r)  = - \, \frac{2}{\varepsilon_0(r) + p_0(r)} \, \frac{d p_0}{dr}(r) \, ,
\label{nr4}
\end{equation}

\noindent with the initial boundary conditions

\begin{eqnarray}
0 < p_0(0) & = & p_{0 c} < \infty \ \mbox{(central pressure),}\nonumber\\
m(0) & = & 0 \, , \ \ \mbox{and} \nonumber\\
\nu(0) & = & \nu_c \ \mbox{(constant fixed by the asymptotic condition at
infinity),}\nonumber
\end{eqnarray}

\noindent this being the prescription for the interior of the star, that is, inside
the fluid, $r \le R,$ \ \hbox{$R \equiv$ radius of} the surface of the star, determined by 
$p_0(R) = 0.$ Furthermore, we assume $p_0$ and $\varepsilon_0$ 
related to each other by a barotropic equation of state,

\begin{equation}
\varepsilon_0 = \varepsilon_0(p_0) \, ,
\label{nr2}
\end{equation}
$p_0 \mapsto \varepsilon_0(p_0)$ a bounded function on any closed interval,
and satisfying the weak energy condition 
\begin{equation}
\varepsilon_0 + p_0 \ge 0 \, .
\label{energycond}
\end{equation}
Observe, from Eqs.~(\ref{nr1}) and (\ref{energycond}), $p_0$ is a decreasing function
from the center, $r=0,$ to the star's surface, $r=R;$ in particular, $p_0 \ge 0$ 
and attains its maximum value $p_{0 c}$ at the center. 

In the exterior (vacuum)
the geometry is described by the same line element (\ref{nrmetric}), but
with the metric function $\nu$ specified and related to $\lambda$ by

\begin{equation}
e^{\nu(r)} = e^{-\lambda(r)} = 1 - \frac{2M}{r},\ \ \forall r > R \, ,
\label{nrext}
\end{equation}

\noindent where $M \equiv m(R)$ is the star's total mass.

Notice that this standard form of the non-rotating metric, (\ref{nrmetric}), is the limit 
of zero rotation of the general rotating metric in spherical polar coordinates (\ref{metric}),
$$
H \to H_0 \equiv  e^{\nu / 2},\ \ Q \to Q_0 \equiv
e^{\lambda / 2},\ \ K \to K_0 \equiv 1, \ \ A \to A_0 \equiv 0 \, ;
$$
\noindent from where the effect of the rotation can be seen as given by the term
$d\phi - A dt$ in the place of $d\phi.$

Note, since, as has been seen above in Subsec.~IV.A, at first order 
in the angular velocity there is still no effect on the pressure, and on the 
energy density, conditions (\ref{nr2}) and (\ref{energycond}) for the starting 
non-rotating configuration are (vi) and (vii) of Subsec.~II.A (at first order).

\subsection*{B. Euler equation}

It will be important to note that at first order
the integrability condition of the Euler
equation and, hence, Eq.~(\ref{Euler_intc}) are no
restriction on $\Omega(r,\theta),$ that shows up in the following.
Consider the first integral of the Euler equation (\ref{Euler_eq}),
namely,

\begin{equation}
\int_0^{p(r,\theta)} \frac{d \bar{p}}{\varepsilon(\bar{p}) +
\bar{p}} + \left.\frac{1}{2} \ln[(u^t)^{-2}]\right|_{(r,\theta)} +
\int_{\Omega_0}^{\Omega(r,\theta)} {\cal F}(\Omega) \, d \Omega =
\mathrm{const.} \, , \label{Euler_1st_int}
\end{equation}

\noindent where Eqs.~(\ref{acc}) and (\ref{Euler_intc}) have been used, and $\Omega_0$ is
a given constant (changing the value of $\Omega_0$ simply modifies the value
of the constant on the right hand side). The first term in Eq.~(\ref{Euler_1st_int})
is a function of the pressure, which is, to this approximation, a
function of $r$, i.e. $O(1)$ with respect to the angular
velocity; on the other hand,
$$
(u^t)^{-2} = -(g_{tt} + 2 \Omega g_{t \phi} + \Omega^2 g_{\phi \phi}) =
H^2 - K^2 r^2 \sin^2 \!\theta \, (\Omega - A)^2 = O(1 - (\Omega -
A)^2) \, ,
$$
\noindent so the second term is $O((\Omega - A)^2)$ and, hence, $O(\mu^2);$ 
also, since
$$
u^t u_{\phi} =(u^t)^2 g_{\phi \phi} (\Omega - A)= \frac{K^2r^2
\sin^2 \!\theta \, (\Omega - A)}{H^2 - K^2 r^2 \sin^2 \!\theta \, (\Omega -
A)^2}= O(\Omega - A) \, ,
$$
\noindent ${\cal F}(\Omega) = u^t u_{\phi} = O(\Omega - A),$ thus, the
third term is $O((\Omega - A)^2)$ as well, and, hence, $O(\mu^2).$
Consequently, to $O(\mu),$ the Euler equation reduces to its
static (non-rotating) case, and indeed we have presumably already used it 
to get the starting unperturbed solution. Therefore, at this order 
in the angular velocity, Eq.~(\ref{Euler_intc}) is no restriction on
$\Omega(r,\theta).$

\subsection*{C. The angular momentum density}

Using the definition of $\omega,$ linear correction of the dragging rate,
via the expansion of the metric coefficient
$g_{t \phi},$ Eq.~(\ref{omegaexp}), and the metric coefficients of
the non-rotating  configuration (\ref{nrmetric}), we obtain the expansion
for the angular momentum density (\ref{amd})

\begin{eqnarray}
T_{\alpha}^{\ \beta} \eta^{\alpha} n_{\beta} & = & n_t T_{\phi}^{\ t} = n_t \, (\varepsilon + p)
(u^t)^2 (g_{t \phi} + \Omega g_{\phi \phi}) \nonumber\\
& = & (n_t)_0 \, (\varepsilon_0 + p_0) [(-g_{tt})_0]^{-1} (- \omega \ (g_{\phi \phi})_0
+ \Omega \ (g_{\phi \phi})_0 ) + O(\mu^3) \nonumber\\
& = & e^{\nu / 2} \, (\varepsilon_0 + p_0) e^{-\nu} \, r^2 \sin^2 \!\theta \ (\Omega - \omega)
+ O(\mu^3) \nonumber\\
& = & (\varepsilon_0 + p_0) \, e^{-\nu / 2} \, r^2 \sin^2 \! \theta \ (\Omega - \omega) + 
O(\mu^3) \, .
\label{amdexp}
\end{eqnarray}

\noindent Thus showing also for the first order rotational perturbation 
that, since we are assuming the energy condition
$\varepsilon_0 + p_0 \ge 0,$ the sign of the angular momentum density (\ref{amdexp})
to $O(\mu)$ is determined by the sign of $\Omega - \omega.$

\subsection*{D. The {\em rotational mass-energy}}

In Ref.~2 Hartle has derived the difference in total mass-energy, $M_{\mathrm{rot}},$ 
between a slowly and differentially rotating relativistic star and a non-rotating star
with the same number of baryons and the same distribution of entropy, namely,
$$
M_{\mathrm{rot}} = {1 \over 2} \int_{\cal I} \Omega \, d J + O(\mu^4) \, ,
$$
where $d J$ is the angular momentum of a fluid element in the star (to first order
in the angular velocity), i.e. from (\ref{Jh}),
$$
d J =  T_{\phi}^{\ t} (- g)^{1/2} d^3 x |_{O(\mu)} \, ;
$$
taking into account (\ref{amdexp}) and (\ref{nrmetric}), we obtain an explicit expression
for the expansion of $M_{\mathrm{rot}}$ in powers of the angular velocity parameter $\mu,$
\begin{equation}
M_{\mathrm{rot}} = {1 \over 2} \int_0^R \negthickspace dr \int_0^{\pi} \negthickspace d \theta \, 
2 \pi (\varepsilon_0 + p_0) \, r^4 \, e^{(\lambda - \nu)/2} \, \sin^3 \!\theta \, 
\Omega (\Omega - \omega) + O(\mu^4) \, .
\label{orMrot}
\end{equation}

\subsection*{E. The time-angle component of the Einstein equation}

The $(t \phi)$ field equation component retaining only first order terms in
the angular velocity, i.e. from Eq.~(\ref{Einstein_eqs}),
$$
R_{\phi}^{\ t} = 8 \pi T_{\phi}^{\ t} + O(\mu^3) \, ,
$$
takes the form
\begin{equation}
\partial_{r} [r^4 j(r) \, \partial_{r} \omega] +
\frac{r^2 \, k(r)}{\sin^3 \!\theta} \partial_{\theta} [\sin^3 \!\theta \, 
\partial_{\theta} \omega] - 16 \pi \, r^4 k(r) [\varepsilon_0(r) + p_0(r)] \,
[\omega - \Omega] = 0 \, , \ \ \
\label{pceq}
\end{equation}

\noindent where we have introduced the abbreviations
\begin{equation}
j(r) \equiv e^{-[\lambda(r) + \nu(r)]/2} \ \ \ \
\mbox{and} \ \ \ \ k(r) \equiv e^{[\lambda(r) - \nu(r)]/2} \, .
\label{jk_def}
\end{equation}
\noindent As outlined in Ref.~1, using the $0$-order field equations,
(\ref{nr0}), (\ref{nr1}), (\ref{nr3}) and (\ref{nr4}), it follows
\begin{equation}
4 \pi \, r \, [\varepsilon_0(r) + p_0(r)] \, k(r) = - j'(r) \, ,
\label{0-ord}
\end{equation}
\noindent (where ${}' \equiv \, d / dr \, )$
which, substituted into Eq.~(\ref{pceq}), yields
\begin{equation}
\partial_{r} [r^4 j(r) \, \partial_{r} \omega] +
\frac{r^2 \, k(r)}{\sin^3 \!\theta} \, \partial_{\theta} [\sin^3 \!\theta \,
\partial_{\theta} \omega]
 + 4 \, r^3 j'(r) \, \omega \
 = 4 \, r^3 j'(r) \, \Omega(r,\theta) \, . 
\label{ceq}
\end{equation}
We write this differential equation for the dragging rate in the
abbreviated form

\begin{equation}
\bar{L} \omega = -\Psi^2 \Omega \, ,
\label{eq}
\end{equation}

\noindent with the linear second order partial differential operator \ $\bar{L}
\equiv \bar{L}_0 - \Psi^2,$
where

\begin{eqnarray}
\bar{L}_0 \omega & := & \frac{1}{r^4 j(r)} \partial_{r} [r^4 j(r) \,
\partial_{r} \omega] +
\frac{k(r)}{r^2 j(r) \sin^3 \!\theta} \, \partial_{\theta} [\sin^3 \!\theta \,
\partial_{\theta} \omega]
\ \ \ \ \mbox{and} \ \ \label{bL_0} \\
\Psi^2(r) & := & - \frac{4}{r} \frac{j'(r)}{j(r)} \ = \
16 \pi [\varepsilon_0(r) + p_0(r)] \, \frac{k(r)}{j(r)} \ge 0 \ \ \ \forall
r \ge 0 \, .
\label{ps2}
\end{eqnarray}

\noindent Equation (\ref{0-ord}) has been used in (\ref{ps2}), and the sign
follows from the assumed energy condition (\ref{energycond}), the
functions $j$ and $k$ (\ref{jk_def}) are always positive.
$\Psi^2 \equiv 0$ in the exterior $(\forall r \in [R,\infty[),$
where vacuum $(\varepsilon_0 = p_0 = 0,$ cf.~(iii) in Subsec.~II.A) is
considered.

\vspace*{2mm}

\noindent Specifically, we are only interested in solutions $\omega \equiv
\omega(r,\theta)$ of
Eq.~(\ref{eq}) in $[0,\infty[ \times [0,\pi],$
which satisfy the {\em boundary conditions}

\begin{eqnarray}
& & \omega \ \ \mbox{asymptotically flat} \ \ 
\left( \lim_{r \to \infty} \omega =  0\right), \ \label{af} \\
& & \omega \ \ \mbox{$C^1$-regular on the axis of rotation,} \ \label{reg}
\end{eqnarray}

\noindent and a {\em matching condition}, namely, to be at least a class $C^1$
function on the surface
of the star --which is spherical at first order rotational perturbations--

\begin{eqnarray}
& & \omega(.,\theta) \ \ \mbox{class} \ \ C^1 \ \ \mbox{across} \ \ r=R \,
. \ \label{mc}
\end{eqnarray}

\noindent Notice, (\ref{af}) follows from our star model (condition (ii) in
Subsec.~II.A),
and it can be easily seen that (\ref{mc}) follows
from the equation itself, provided that $\omega(.,\theta)$ and $\Omega(.,\theta)$ 
are at least essentially bounded $(\in L^{\infty})$ ---as has been assumed---
i.e. even if they have a jump discontinuity. 

At the star's surface, $r=R,$ higher regularity of $\omega(.,\theta)$ is
not guaranteed by
the equation, due to a jump discontinuity of the function $\Psi^2$ at this
point. For this reason, we shall be considering (in the following section)
{\em generalized} $(\in W^{1,2})$ solutions $\omega$ 
of Eq.~(\ref{eq}) in the whole domain (interior and exterior). 

\bigskip

\vspace*{1mm}

\noindent{\large{{\bf ``Coordinate change''.}}}

\vspace*{1mm}

In order to avoid the coordinate singularity occurring on the axis in
polar coordinates $(r,\theta),$ and wishing instead to have in the 
differential operator (\ref{bL_0}) a Lapacian in some higher dimension, we
consider the following ``change of coordinates''.
Firstly, we introduce
{\em isotropic} cylindrical coordinates in the meridian plane,
\begin{equation}
(r,\theta) \mapsto (\, \rho:= h(r) \sin\theta \ , \ z:= h(r) \cos\theta \, ) 
\ \in \R^+_0 \times \R \, ,
\label{ch1}
\end{equation}
with the function $h$ satisfying the following ordinary differential equation 
of first order with separated coefficients
\begin{equation}
\frac{h'(r)}{h(r)} = \frac{e^{\lambda(r) / 2}}{r}
\, 
\label{eq_h}
\end{equation}
(which makes the coefficient of the crossed derivatives in the operator (\ref{bL_0}) 
after the change (\ref{ch1}) to vanish),
and the {\em boundary condition} 
\begin{equation}
\lim_{r \to \infty} \frac{h(r)}{r} = 1 \, ,
\label{has}
\end{equation}
i.e. so that the {\em isotropic} radius \ $h(r) \equiv \bar{r}$ \ approaches \ $r$ \ 
at spatial infinity, because far away from the source we assume to have 
euclidean geometry. This leads us to the definition of the function 
\begin{equation}
w(\rho,z):= \omega \left(h^{-1}\left(\sqrt{\rho^2 + z^2}\right) \, , 
\, \arctan \left( \frac{\rho}{z} \right)\right) 
\, ,
\label{w}
\end{equation}
or inversely, \/ $w$ \/ such that
$$
\omega(r,\theta) = w(h(r) \sin\theta \, , \, h(r) \cos\theta) \, .
$$
Secondly, (in the spirit of Ref.~10) we define (with $w(\rho,z))$ 
the $5$-{\em lift of}
$w: \R^+_0 \!\times \! \R \rightarrow \R$ in flat $\R^5,$ axisymmetric around the
$x_5$-axis, by
\begin{equation}
w \mapsto \tilde{\omega} \quad \mbox{such that} \quad 
\tilde{\omega}({\bf x}) \equiv \tilde{\omega}(x_1,x_2,x_3,x_4,x_5):=
w\left(\rho = \left(\sum_{i=1}^4x_i^{\ 2}\right)^{1/2} \ , \ z = x_5 \right) \, ,
\label{ch2}
\end{equation}
and, for every function $\tilde{\omega}: \R^5 \rightarrow \R,$ 
the {\em meridional cut (in direction $x_1$) of} $\tilde{\omega}$
$$
\tilde{\omega} \mapsto w \quad \mbox{such that} \quad 
w(\rho,z):= \tilde{\omega}(\rho,0,0,0,z) \, .
$$
For axisymmetric functions these are isometric operations inverse to each other \cite{Sch}.
After considering the change of variable (\ref{ch1}) with (\ref{w})
in the differential operator $\bar{L}_0$
(\ref{bL_0}), we get (remember (\ref{jk_def}), $e^\lambda = k/j)$

\begin{equation}
\bar{L}_0 \, w
 = 
 \frac{e^{\lambda(r)}h(r)^2}{r^2} 
 \Big\{ \partial_{\rho \rho} w + \partial_{zz} w + \frac{3}{\rho}\, 
\partial_{\rho} w
+ H(r) \frac{\rho\, \partial_{\rho} w + z \, \partial_{z} w}{h(r)} \Big\} \, ,
\label{nL_0}
\end{equation}

\noindent where 
\begin{equation}
    H(r) :=\frac{-e^{\frac{-\lambda (r)}{2}}\,
          [-6 + 6\,e^{\frac{\lambda (r)}{2}} + r\,\nu '(r)] 
           }{2\,h(r)} \, .
\label{H}
\end{equation}
But, through the $5$-lift (\ref{ch2}),
the flat Laplacian in $5$ dimensions of the ``lifted'' function $\tilde{\omega}$ gives exactly
\begin{equation}
\Delta \tilde{\omega} \, \equiv \, \sum_{i=1}^5\partial_{ii} \tilde{\omega} \, = \, 
\partial_{\rho \rho} w + \partial_{zz} w + \frac{3}{\rho}\, \partial_{\rho} w \, ,
\label{laplacian}
\end{equation}
first three terms in the bracket of (\ref{nL_0}).
Furthermore, as outlined in Ref.~10, $n$-lift and meridional cut 
(of {\em axisymmetric functions})
leave the regularity properties and the norm invariant; and {\em axisymmetric} operations, like
multiplication, $\partial_{\bar{r}}$ \ $(\bar{r} \equiv h(r) = 
(\rho^2 + z^2)^{1/2} = (\sum_{i=1}^5 x_i^{\, 2})^{1/2} \, ),$ and scalar 
product, commute with $n$-lift and meridional cut. In particular, in the fourth term in the bracket 
of (\ref{nL_0}) the factor
\begin{equation}
\rho\, \partial_{\rho} w + z \, \partial_{z} w \, = \, \partial_{\bar{r}} \tilde{\omega} \, = \,
\sum_{i=1}^5 x_i \, \partial_i \tilde{\omega} \, .
\label{product}
\end{equation}
Therefore, substituting (\ref{laplacian}) and (\ref{product}) into (\ref{nL_0}),
Eq.~(\ref{eq}) in the form
$\bar{L}_0 \tilde{\omega} = - \Psi^2 (\tilde{\Omega} - \tilde{\omega})$ \  
(with $\tilde{\Omega}$ defined from $\Omega$ as it was $\tilde{\omega}$ from $\omega,$
and $\Psi^2 = e^{\lambda} 16 \pi [\varepsilon_0 + p_0] \, )$ 
now writes
$$
\bar{L}_0 \, \tilde{\omega}
 \equiv
 \frac{e^{\lambda(r)}h(r)^2}{r^2} 
 \Big\{ \Delta \tilde{\omega} 
+ H(r) \frac{ \sum_{i=1}^5 x_i \, \partial_i \tilde{\omega}}{h(r)} \Big\} 
= e^{\lambda(r)}\left\{ -16 \pi [\varepsilon_0(r) + p_0(r)] [\tilde{\Omega} - \tilde{\omega}]
\right\} \, ,
$$
or, equivalently \cite{fn},

\begin{equation}
\Delta \tilde{\omega} 
+ H(r) \frac{\sum_{i=1}^5 x_i \, \partial_i \tilde{\omega}}{h(r)} 
= -16 \pi \frac{r^2}{h(r)^2} [\varepsilon_0(r) + p_0(r)]  [\tilde{\Omega} - \tilde{\omega}] \, .
\label{peq}
\end{equation}

\section*{\normalsize V. PROPERTIES}

With the assumptions made in Subsec.~II.A for this slowly rotating
configuration (starting from
a non-rotating one as described in Subsec.~IV.A.1; particularly,
satisfying the energy condition $\varepsilon_0 + p_0 \ge 0$ ), 
and considering only solutions of Eq.~(\ref{eq}) satisfying 
the boundary and matching conditions (\ref{af}), (\ref{reg}), and (\ref{mc}),
the following results hold

\vspace*{2mm}

\noindent {\bf Property (a) (Positiveness of the {\em dragging rate})}

\vspace*{1mm}

\noindent {\it If the distribution of angular velocity of the
fluid is non-negative (and non-trivial), then the {\em dragging rate} (to first order in the
fluid angular velocity) is positive everywhere,}
$$
\Omega \ge 0, \ \ \Omega \not \equiv 0 \ \ \
\Longrightarrow \ \
\omega > 0 \, .
$$

\vspace*{2mm}

\noindent {\it Proof.}
We have seen in the former section that Eq.~(\ref{eq}) for $\omega$ is equivalent to 
Eq.~(\ref{peq}) for $\tilde{\omega}$ using the coordinate change (\ref{ch1}) and the
$5$-lift (\ref{ch2}). The isotropic radius $\bar{r} \equiv h(r)$ is Gaussian coordinate
with respect to the star's surface, $\bar{r} = h(R),$ and, thus, $\tilde{\omega}$ is
at least class $C^1$ across this surface; therefore, from conditions on 
the stellar model, $\tilde{\omega} \in W^{1,2}(\R^5) \cap C^{1}(\R^5),$ and, hence,
Eq.~(\ref{peq}), i.e. 
\begin{equation}
L \tilde{\omega} := \Delta \tilde{\omega} 
+ \sum_{i=1}^5 \, H(r) \frac{x_i}{h(r)} \, \partial_i \tilde{\omega}
- 16 \pi [\varepsilon_0 + p_0] \frac{r^2}{h(r)^2} \tilde{\omega}
= - 16 \pi [\varepsilon_0 + p_0] \frac{r^2}{h(r)^2} \tilde{\Omega} \, ,
\label{equ}
\end{equation}
is satisfied in $\R^5$ in a generalized sense (cf.~Appendix B).
Equation~(\ref{equ}) may be obviously written {\em in divergence form}
$$
L \tilde{\omega} \equiv \partial_{i} [a^{ij}({\bf x}) \partial_{j} \tilde{\omega} 
+ a^i({\bf x}) \, \tilde{\omega}]
+ b^i({\bf x}) \, \partial_{i} \tilde{\omega} + c({\bf x}) \, \tilde{\omega} \, = \, g({\bf x}) 
$$
(where repeated indices denote summation over the index),
with the coefficients
\begin{eqnarray}
a^{ij}({\bf x}) & \equiv & \delta_{ij} \ \ (= 1 \ \mbox{if} \ i=j, \ \mbox{and} \ = 0 \ 
\mbox{otherwise}) \, , \nonumber\\
a^i({\bf x}) & \equiv & 0 \, ,\nonumber\\
b^i({\bf x}) & = & H(r) \frac{x_i}{h(r)} \ \ \ \ \
(\forall \, i,j \in \{1,\ldots, 5\}), \ \ \mbox{and}  \label{cf} \\ 
c({\bf x}) & = & - 16 \pi [\varepsilon_0(r) + p_0(r)] \frac{r^2}{h(r)^2} \ \ (\le 0) \, , \nonumber\\
\mbox{and} \ \ g({\bf x}) & = & c({\bf x}) \, \tilde{\Omega}({\bf x}) \, . 
\label{g}
\end{eqnarray}
We shall consider the domain $G$ defined by a ball \/ in $\R^5$ centered at the origin \ 
${\bf x} = {\bf 0} \ (\bar{r} = 0)$ and of arbitrary large radius $\sigma,$ \ 
\begin{equation}
G := {\cal B}_ {\sigma} ({\bf 0}) \, \subset \R^5 \, .
\label{B}
\end{equation}
Notice, whenever $\tilde{\Omega} \ge 0,$ we have, by (\ref{g}), $g \le 0$ (because $c \le 0),$
and, hence, $L \tilde{\omega} \le 0,$ specifically $\tilde{\omega} \in W^{1,2}(G) \cap C^1(G)$ 
is a generalized supersolution relative to the operator $L,$ in (\ref{equ}),
and the domain $G,$ (\ref{B}). We look at the requirements for a minimum principle to be
applied (Appendix B). The Laplacian operator is obviously
strictly elliptic, and the coefficients (\ref{cf}) are measurable and bounded
functions on $G,$ this shows up in the following:
the mapping $r \mapsto \frac{r}{h(r)}$ is bounded from above and below everywhere in $[0,\infty[$
(Appendix A); $\varepsilon_0 + p_0$ is also bounded, since $p_0$ is bounded and 
$p_0 \mapsto \varepsilon_0(p_0)$ is bounded in any closed interval (Subsec.~IV.A.1);
consequently, the coefficient $c$ is bounded 
(from above and below); the coefficients of the first 
derivatives, $b^i,$ are also bounded (from above and below), because the function $H$ 
is bounded everywhere (Appendix A) and since $(\forall i = 1,\ldots,5)$ \ 
$x_i^{\ 2} \le \sum_{j=1}^{5} x_j^{\ 2} = h(r)^2,$ 
we have $(x_i / h(r))^2 \le 1.$ 
Thus, all conditions of a minimum principle for generalized supersolutions
relative to the differential operator $L$ and the domain $G$ hold, and, as a result of
the {\em weak} minimum principle (Theorem 1 in Appendix B), we have
\begin{equation}
\inf_G \tilde{\omega} \ge \inf_{\partial G} \tilde{\omega}^{-}, \ \ \ 
(\tilde{\omega}^{-} \equiv \min(\tilde{\omega},0) \, ) \, .
\label{res}
\end{equation}
But, since the radius of the ball $G$ is arbitrary, we can make it sufficiently large 
$(\sigma \to \infty)$ 
so that, by asymptotic flatness $( \, \lim_{h(r) \to \infty} \tilde{\omega} = 0,$
from condition (\ref{af}) and $\lim_{r \to \infty} \frac{h(r)}{r} = 1),$
\ $\tilde{\omega}$ is arbitrary small
at $\partial G,$ following, from (\ref{res}), $\tilde{\omega} \ge 0.$ Actually, the
positivity is strict, because if $\tilde{\omega}({\bf x_0}) = 0$ for some ${\bf x_0} \in G$
(interior point), then $\tilde{\omega}({\bf x_0}) = \min \, \tilde{\omega}$ 
(since $\tilde{\omega} \ge 0),$ and, by
the {\em strong} minimum principle (Theorem 2 in Appendix B), $\tilde{\omega}$ would be constant
in $G;$ in this case, $\tilde{\omega} \equiv$ const. $= 0$ in $G$ (i.e. everywhere); but
$\tilde{\omega} \equiv 0$ yields, by Eq.~(\ref{equ}), $\tilde{\Omega} \equiv 0,$ or, equivalently,
$\Omega \equiv 0,$ and we are assuming that $\Omega$ is non-trivial. 
We conclude then $\omega > 0$ everywhere. \hspace*{\fill}$\square$

\vspace*{4mm}


\noindent {\bf Property (b)}

\vspace*{1mm} 

\noindent {\it Suppose we perturb the non-rotating 
configuration (in particular, with a given equation of state) with 
two (small) different distributions of angular velocity
$\Omega_1$ and $\Omega_2,$ and integrate Eq.~{\em (\ref{eq})}
to obtain their respective solutions for the dragging rate,
$\omega_1$ and $\omega_2,$ then}\\
$$
\Omega_1 \ge \Omega_2, \ \ \ \Omega_1 \not \equiv \Omega_2 \ \ \
\Longrightarrow \ \
\omega_1 > \omega_2 \  .
$$
 
\vspace*{2mm} 

\noindent {\it Proof.} \ This follows form the linearity of Eq.~(\ref{eq})
and Property (a). \hspace*{\fill}$\square$

\vspace*{5mm} 


We are already in position to get a result about the positiveness of the 
difference $\Omega - \omega,$ and, hence, of the {\em angular momentum density}
(\ref{amdexp}). However, in order to first do this more specific and concrete, we shall make
use of a property for the particular case of rigid rotation (RR), which can
be found in Ref.~1, Sec.~IV. 

\vspace*{2mm}

\noindent {\bf Property RR}

\noindent {\it For the slowly rotating configuration,
\begin{eqnarray}
& & \omega(r,\theta) = \omega(r)\nonumber\\
& \Omega(r,\theta) \equiv \ \mbox{{\em const.}}
\equiv \widehat{\Omega} > 0 \ 
\Longrightarrow & 
0 < \omega(r) < \widehat{\Omega} \ \ \mbox{in} \ [0,R] \nonumber \\
& ({\mbox in} \, [0,R] \! \times \! [0,\pi] \, ) \ &  
\omega > 0 \, \mbox{in} \ [0,\infty[ \, , \ \omega' < 0 \ \mbox{in} \ ]0,\infty[ \, ,
\ \omega'(0) = 0 \, .
 \nonumber
\end{eqnarray}
}

\noindent {\bf Property (c) (Positiveness of the {\em angular momentum density})}

\vspace*{1mm} 

\noindent {\it For the slowly rotating configuration, with a given equation of state,
its {\em dragging rate} $\omega$ will satisfy 
$$
\omega(r,\theta) < \Omega(r,\theta)
$$
\noindent if $\Omega \equiv \Omega(r,\theta) \ ( \ge 0)$ is bounded in the form
$$
\underline{\Omega}(\overline{\Omega}) \equiv \underline{\Omega} 
\ \le \ \Omega(r,\theta) \ \le \ \overline{\Omega} \, ,
$$
(in $[0,R] \times [0,\pi]$) where $\overline{\Omega}$ is an arbitrary positive constant 
$0 < \overline{\Omega} \ ( \ll \Omega_{crit} )$ (Sec.~IV), and 
$\underline{\Omega} = \overline{\omega}(0),$ \ $\overline{\omega}$ solution of Eq.~{\em (\ref{eq})}
with $\overline{\Omega}(r,\theta) =$ {\em const.} $= \overline{\Omega},$ and with the same 
$0$-order coefficients (same starting non-rotating configuration) as the ones considered
for our slowly rotating configuration, in particular with the same equation of state.

\vspace{2mm}

\noindent Or, more generally, if (with that notation)
$$
\overline{\omega}(r)
\ \le \ \Omega(r,\theta) \ \le \ \overline{\Omega} \, .
$$
Notice, $\omega < \Omega$ means that the {\em angular momentum density}
to first order in the fluid angular velocity, {\em (\ref{amdexp})}, of this configuration
(with the energy condition {\em (\ref{energycond})} \ ) is $\ge 0,$ vanishing on the axis.
}

\vspace{2mm}

\noindent (Remarkably, the upper bound required on $\Omega$ is not restrictive, 
because for $\Omega$ continuous, $\Omega$ is essentially bounded 
$(\in L^{\infty})$ there, and $\Omega / \| \Omega \|_{\infty} \, \le 1.$)

\vspace*{2mm}

\noindent {\it Proof.} 
We give a practical method of construction in two steps:
\begin{description}
\item[1st step:] Consider $\overline{\Omega}(r,\theta) := \overline{\Omega} =$ const. $>0,$
and solve the corresponding Eq.~(\ref{eq}) for $\overline{\omega}.$
Then, by Property RR, the solution satisfies 

\begin{eqnarray}
& & \overline{\omega}(r,\theta) = \overline{\omega}(r) \, ,\label{rRR}\\
& & 0 < \overline{\omega}(r) < \overline{\Omega} \ \ \mbox{in} \ [0,R] \, ,\label{bRR} \\
& &  \mbox{and} \ \ \overline{\omega} > 0 \, \mbox{in} \ [0,\infty[ \, , \ \ \overline{\omega}' < 0 
\ \mbox{in} \ ]0,\infty[ \, , \ \ \overline{\omega}'(0) = 0 \, . \label{behRR}
\end{eqnarray}

\item[2nd step:] Consider a slowly rotating configuration starting from the same
non-rotating configuration (as in the first step) with a fluid angular velocity 
distribution $\Omega(r,\theta)$ such that
\begin{equation}
\overline{\omega}(0) =: \underline{\Omega} \ \le \ \Omega(r,\theta) \ \le \ \overline{\Omega} \, .
\label{ineq}
\end{equation}
Observe, we are always allowed to do this because of (\ref{bRR}).
Or, more generally, such that 
$ \overline{\omega}(r) \le \Omega(r,\theta) \le \overline{\Omega};$ notice,
from (\ref{behRR}), $\overline{\omega}$ is a (positive) 
decreasing function; in particular, $\overline{\omega}(0) \ge \overline{\omega}(r) > 0,
\ \ \forall r \in [0,\infty[.$

\vspace{1mm}

\noindent Now, form the second inequality in (\ref{ineq}), i.e. from 
$\Omega(r,\theta) \le \overline{\Omega},$ and, since we have the
same starting unperturbed configuration (same 0-oder coefficients)
for these both slowly rotating configurations,
it follows, by Property (b), that their corresponding solutions (of Eq.~(\ref{eq}))
satisfy
\begin{equation}
\omega(r,\theta) < \overline{\omega}(r) \, ,
\label{in1}
\end{equation}
where we have used (\ref{rRR}).
On the other hand, the first inequality in (\ref{ineq}), and (\ref{behRR}) yield
\begin{equation}
\overline{\omega}(r) \le \overline{\omega}(0) := \underline{\Omega} \ \le \ \Omega(r,\theta) \, ,
\label{in2}
\end{equation}
and consequently, from (\ref{in1}) and (\ref{in2}), 
$$
\omega(r,\theta)  \ < \ \Omega(r,\theta) \, .
$$
\hbox{}\hfill$\square$
\end{description}

\noindent {\it Remark.}
Notice, the same argument also assures that, given a slowly rotating configuration with 
$\overline{\Omega}(r,\theta)$ such that
the corresponding dragging rate $\overline{\omega}(r,\theta) < \overline{\Omega}(r,\theta),$ 
we shall have the same positivity result (Property (c))
for any slowly rotating configuration, starting
from the same unperturbed configuration (in particular, with the same equation of state),
with an angular velocity distribution $\Omega(r,\theta)$ such that
$$
\underline{\Omega}(r,\theta) := \overline{\omega}(r,\theta) \ \le \ \Omega(r,\theta) 
\ \le \ \overline{\Omega}(r,\theta) \, ,
$$
because we obtain, form the last inequality and Property (b), 
$\omega(r,\theta) < \overline{\omega}(r,\theta),$ and, hence, 
$\omega(r,\theta) < \Omega(r,\theta).$

\bigskip

\vspace*{1mm}
\noindent{\large{{\bf Series expansion. \ $M_{\mathrm{rot}}$.}}}

\vspace*{1mm}

Before we prove next property, we first stress that, since $\Omega$ and $\omega$
transform like vectors under rotation, Eq.~(\ref{ceq}) may be separated
by expanding them as

\begin{eqnarray}
\Omega(r,\theta) & \equiv & \Omega(r,x) \sim \sum_{l=1}^{\infty} \Omega_l(r) \, y_l(x) 
\ \ \mbox{and} \label{Omexp}\\
\omega(r,\theta) & \equiv & \omega(r,x) \sim \sum_{l=1}^{\infty} \omega_l(r) \, y_l(x) \, ,
\label{omexp}
\end{eqnarray}

\noindent with the change of variable $\theta \mapsto x := \cos\theta,$ \ and  
\begin{equation}
y_l(x) := \frac{d \mathcal{P}_l}{dx} 
\ \ \ \forall x \in [-1,1] \ \ (\theta \in [0,\pi]), \ \ \ \mathcal{P}_l \equiv
\mbox{Legendre polynomial of degree} \ l \, .                                                      
\label{y_l}          
\end{equation}       
\noindent Then the equation for $\omega_l$ takes the form                                           
                                                                        
\begin{equation}
\frac{d}{d r} [ r^4 j(r) \ \omega_l' ] +
[ 4 \, r^3 j'(r) - r^2 \, k(r) \, \lambda_l ] \ \omega_l =
4 \, r^3 j'(r) \ \Omega_l(r) \, ,
\label{preq}
\end{equation}
\noindent with $\lambda_l := l(l+1) - 2, \ \ l \in \N,\ l \neq 0,$ and
$j$ and $k$ defined in (\ref{jk_def}).

\noindent From conditions (\ref{af}) and (\ref{reg}) on $\omega,$ we have 
the respective boundary conditions on $\omega_l$
\begin{eqnarray}
& & \lim_{r \to \infty} \omega_l(r) =  0 \, , \label{laf} \\
& & \omega_l \ \ \mbox{$C^1$-regular at the origin}, \label{lreg} 
\end{eqnarray}
\noindent and, from (\ref{mc}), the matching condition
\begin{eqnarray}
& & \omega_l \ \ \mbox{class} \ \ C^1 \ \ \mbox{across} \ \ r=R \,
. \ \label{lmc}
\end{eqnarray}

In Subsec.~IV.D an explicit expression for the expansion of the rotational mass-energy
$M_{\mathrm{rot}}$ in powers or the angular velocity parameter was obtained (\ref{orMrot}), 
or, using Eq.~(\ref{0-ord}), 
\begin{equation}
M_{\mathrm{rot}} = - \, {1 \over 4} \int_0^R dr \, r^3  \, j' \int_0^{\pi} d \theta 
\sin^3 \!\theta \, \Omega (\Omega - \omega) + O(\mu^4) \, .
\label{Mrot}
\end{equation}

\noindent Now, using the series expansions of $\Omega$ and $\omega$ 
(\ref{Omexp}) and (\ref{omexp}), and
the fact that the system $\{y_l\}_{l=1}^{\infty}$ is orthogonal in the Hilbert 
space $L_{\rho}^2 ([-1,1]),$ with respect to the weight function
$\rho(x) := 1 - x^2, \ x \in [-1,1],$ and have norm 
$\| y_l \|_{\rho}^{\ 2} = 2l(l+1)/(2l + 1),$
the integral over $\theta$ in (\ref{Mrot}) may be expressed as the sum
\begin{equation}
\int_0^{\pi} d \theta \sin^3 \!\theta \, 
\Omega(r,\theta) \left[\Omega(r,\theta) - \omega(r,\theta) \right]
= \sum_{l=1}^{\infty} \frac{2l(l+1)}{2l+1} \Omega_l(r) \left[ \Omega_l(r) - \omega_l(r) \right] \, ,
\label{prev}
\end{equation}
\noindent and, consequently, the rotational mass-energy (\ref{Mrot}) can be expressed as
a sum of integrals (over $r)$
\begin{equation}
M_{\mathrm{rot}} = \sum_{l=1}^{\infty} \frac{l(l+1)}{2(2l+1)} \, M_l \, + \, O(\mu^4) \, ,
\label{Mrot_sum}
\end{equation}
\begin{equation}
\mbox{with} \ \ M_l := \int_0^R f^2(r) \, \Omega_l(r) 
\left[ \Omega_l(r) - \omega_l(r) \right] \, dr \, , \ \ 
\ \ \ \ \ f^2(r):= -r^3 \, j'(r) \ (\ge 0) \, .
\label{M_l}
\end{equation}

\vspace*{4mm}

\noindent {\bf Property (d) (Positivity and upper bound on the {\em rotational energy} $M_{\mathrm{rot}}$)}

\vspace*{2mm}

We consider Eq.~(\ref{preq}), which can be written
\begin{equation}
\frac{d}{d r} (r^4 \, j \, \omega_l')  - r^2 \, k \, \lambda_l \, \omega_l =
- 4 \, f^2 \, (\Omega_l - \omega_l) \, .
\label{req}
\end{equation}
\noindent The main observation is that, multiplying both sides of Eq.~(\ref{req}) by
$\omega_l$ and integrating from $r=0$ to $r=\infty,$ and taking into account that
$f^2 =  -r^3 \, j' =  4 \pi \, r^4 \, (\varepsilon_0 + p_0 ) \, k \equiv 0 \ \ 
\forall r > R,$
$$
\int_0^{\infty} \left[ \frac{d}{d r} (r^4 \, j \, \omega_l') \, \omega_l  
- r^2 \, k \, \lambda_l \, \omega_l^{\ 2} \right] dr =
- 4 \int_0^R f^2 \, \omega_l \, (\Omega_l - \omega_l) \, dr
$$
\noindent (note, the integral on the left hand side converges, because
an asymptotically flat (\ref{laf}) solution of Eq.~(\ref{preq}) must behave
as $r \to \infty$ \ \ $\omega_l = O( r^{-l-2}),$ and $\omega_l' = O (r^{-l-3}),$ $l \ge 1);$
and, after integrating once by parts the first term on the left hand side,
$$
\left. r^4 \, j \, \omega_l' \, \omega_l \right|_0^{\infty} 
- \int_0^{\infty} \left[r^4 \, j \, (\omega_l')^2 
+  r^2 \, k \, \lambda_l \, \omega_l^{\ 2} \right] dr =
- 4 \int_0^R f^2 \, \omega_l \, (\Omega_l - \omega_l) \, dr \, .
$$
Now, the first term vanishes because $\omega_l$ falls off rapidly enough
at $r \to \infty,$ and the second term (minus the integral on the left hand side) 
is non-positive (since $j$ and $k$ are always positive), therefore
\begin{equation}
\int_0^R f^2 \, \omega_l \, (\Omega_l - \omega_l) \, dr \, \ge 0 \, .
\label{r1}
\end{equation}
 
We introduce now the sesquilinear (indeed bilinear) form
$$
\langle u,v \rangle_f := \int_0^R f^2(r) \ u(r) \ v(r) \ dr \ \ \ \ \ \ u,v \in C^0([0,R]),
$$
and the induced semi-norm $\|.\|_f := \left( \langle . \, , . \rangle_f \right)^{1/2}.$ 
With this definition, $M_l$ (\ref{M_l}) may be written
\begin{equation}
M_l = \langle \Omega_l , \Omega_l - \omega_l \rangle_f \, .
\label{sM_l}
\end{equation}
\noindent We have (\ref{r1}), which now reads 
$\langle \omega_l , \Omega_l - \omega_l \rangle_f \ge 0,$ i.e. 
\begin{equation}
\langle \omega_l , \Omega_l \rangle_f \ge  \|\omega_l\|_f^{\ 2} \, ,
\label{r2}
\end{equation}
\noindent in particular
\begin{equation}
\langle \Omega_l , \omega_l \rangle_f = \langle \omega_l , \Omega_l \rangle_f \ge  0  \, .
\label{r3}
\end{equation}
\noindent Using the Cauchy-Schwarz inequality,
$$
\langle \omega_l , \Omega_l \rangle_f \le \|\omega_l\|_f \|\Omega_l\|_f \, ,
$$
which, together with (\ref{r2}), yields $\|\omega_l\|_f^{\ 2} \le \|\omega_l\|_f \|\Omega_l\|_f$
and, hence, $\|\Omega_l\|_f \ge \|\omega_l\|_f$ for $\|\omega_l\|_f \neq 0,$ but this inequality
is still true for  $\|\omega_l\|_f = 0,$ because then, by Eq.~(\ref{preq}), $\|\Omega_l\|_f = 0;$
therefore 
\begin{equation}
\|\Omega_l\|_f \ge \|\omega_l\|_f
\label{r4}
\end{equation}
\noindent holds in general. Multiplying (\ref{r4}) by $\|\Omega_l\|_f,$ we find, again
using the Cauchy-Schwarz inequality,
$$
\|\Omega_l\|_f^{\ 2} \ge \langle \Omega_l , \omega_l \rangle_f \, ,
$$
but
\begin{equation}
M_l = \langle \Omega_l , \Omega_l - \omega_l \rangle_f =
\|\Omega_l\|_f^{\ 2} - \langle \Omega_l , \omega_l \rangle_f \, ,
\label{r5}
\end{equation}
\noindent thus showing
\begin{equation}
M_l \ge 0 \, .
\label{lb}
\end{equation}
\noindent Furthermore, since (by (\ref{r3})) $\langle \Omega_l , \omega_l \rangle_f \ge 0,$
from (\ref{r5}) we also have the upper bound
$$
M_l \le \|\Omega_l\|_f^{\ 2} \, , \ \ \mbox{i.e.}
$$
\begin{equation}
M_l \le \int_0^R f^2(r) \ \Omega_l^{\ 2}(r) \, dr \, .
\label{ub}
\end{equation}
The bounds (\ref{lb}) and (\ref{ub}) on $M_l$ yield respective bounds on
$M_{\mathrm{rot}}$ (\ref{Mrot_sum}),
$$
0 \le M_{\mathrm{rot}} \le \sum_{l=1}^{\infty} \frac{l(l+1)}{2(2l+1)} 
\int_0^R f^2(r) \ \Omega_l^{\ 2}(r) \, dr \, + \, O(\mu^4) \, ,
$$
or, writing the sum as integral over $\theta$ (as in (\ref{prev})),
\begin{equation}
0 \le M_{\mathrm{rot}} \le 
{1 \over 4} \int_0^R dr \, f^2(r) \int_0^{\pi} d \theta \, \sin^3 \!\theta \, 
[\Omega(r,\theta)]^{2} \, + \, O(\mu^4) \, ,
\label{bounds}
\end{equation}
\noindent where
$f^2 := -r^3 \, j' =  4 \pi \, r^4 \, (\varepsilon_0 + p_0 ) e^{(\lambda - \nu)/2}.$
\hbox{}\hfill$\square$

\section*{\normalsize VI. CONCLUDING REMARKS}

Summing up, it has been seen that 
relativistic stars rotating slowly and differentially,
with a non-negative (and non-trivial) angular velocity distribution,
$\Omega(x_2,x_3) \ge 0 \ (\not\equiv 0),$
and satisfying the energy condition $\varepsilon + p \ge 0,$ have 
positive {\em rate of rotational dragging} $\omega > 0$ (Property (a) in Sec.~V);
and certain upper and lower bounds on $\Omega$ assure
also the positivity of the difference $\Omega - \omega$ and,
hence, of the {\em angular momentum density}, this later 
vanishing on the axis, (Property (c)).
We also observe that, 
the {\em rotational mass-energy}, (from Property (d))
non-negative and (as expected) ``increased'' by a
(slow) angular velocity of the fluid, $\Omega,$ is
``decreased'' by the dragging effect (over what it would be is
this effect were neglected), i.e. is decreasing with respect to 
dragging rate, $\omega,$ despite of (as shown in Property (b)) $\omega$ 
being an ``increasing function'' of $\Omega.$ 

We have given here a much simpler proof of Property (d) than in
Ref.~2; this alternative proof can be even generalized
outside the slow rotation limit \cite{dr}. Property (b) and,
hence, also Property (c), however are based on the linearity of
the time-angle field equation component to first order in the
fluid angular velocity. Moreover, in the general differentially
rotating case the rotation profile $\Omega$ cannot be freely
chosen, but is restricted by the integrability condition of the
Euler equation, i.e.~by Eq.~(\ref{Euler_intc}). This makes
unlikely a generalization of Property (c) outside the slow rotation
limit, other than in the form given in Ref.~12, Subsec.~IV.B.

\section*{\normalsize ACKNOWLEDGMENT}
I would like to thank F.J. Chinea and U.M. Schaudt for reading the 
manuscript and for many valuable discussions.
This work was supported by Direcci\'on General de Ense\~nanza Superior,
Project No. PB98-0772.

\appendix
\setcounter{equation}{0}
\renewcommand{\theequation}{A \arabic{equation}}

\section*{\normalsize APPENDIX A: \ Boundedness of some functions 
in $[0,\infty[ \, \ni r$}

\subsection*{\normalsize \ The ratio {\em radius - isotropic radius} \ 
$\chi(r):=\frac{r}{h(r)}$}

\bigskip

\noindent
We have
\begin{equation}\label{ode_h}    
\frac{h'(r)}{h(r)} = \frac{e^{\lambda(r)/2}}{r}    
\,,\end{equation}
with
\begin{equation}\label{lambda}
e^{-\lambda(r)} = 1-\frac{2 \, m(r)}{r}
\,,
\end{equation}
where
\begin{equation}\label{m}    
m(r):=\left\{{    
\begin{array}{lcl}      
4\pi\int_0^r \varepsilon_0(s) s^2\, ds & : & r\in [0,R] \\      
M \equiv 4\pi\int_0^{R} \varepsilon_0(s) s^2\, ds& : & r\in ]R,\infty[ \\    
\end{array}
}    
\right.    
\, ,
\end{equation}
if we denote the stellar radius of the static model by $R>0.$ As we start 
with a (physically) regular (i.e. non-collapsed) static solution, we assume
that $2m(r) < r$ (for all $r\in ]0,R[ \, ),$ and $2M < R.$
\bigskip

\noindent
Integrating Eq.~(\ref{ode_h}) and using Eq.~(\ref{lambda}) we get
\begin{equation}\label{h} 
h(r) = h(R) \, \exp\left(\int_{R}^r\frac{ds}{\sqrt{s(s-2m(s))}}\right)    
\,.
\end{equation}
(Note, the constant $h(R)$ is determined by the asymptotic condition
\[
\lim_{r\to\infty}\frac{h(r)}{r}=1 \,;
\]
see below.)
Let
\begin{equation}\label{def_g}    
g(r) := \int_{R}^r\frac{ds}{\sqrt{s(s-2m(s))}}    
\quad\forall r>0    
\,.
\end{equation}
With this definition the solution, (\ref{h}), now writes
\begin{equation}\label{sol_h}
h(r) = h(R) \, \exp \left( g(r) \right)    
\,.
\end{equation}
Due to the assumptions made for $m,$ $g$ is obviously a continuous function in the 
open interval $]0,\infty[;$ consequently, by (\ref{sol_h}), $h$ is also a continuous
function there, and, in particular, $h(r)$ cannot be zero 
in $]0,\infty[$ (unless $h(R)=0,$ however this would contradict 
asymptotic flatness); therefore $\chi:r \mapsto \chi(r):=\frac{r}{h(r)}$ 
is continuous in $]0,\infty[$ as well. Choose an $\epsilon \in ]0, R[$ and an
$\epsilon' \in ]R,\infty[,$ then $\chi$ is bounded below 
and above in the interval $[\epsilon,\epsilon']$ (where the upper and lower 
bound depend on the selected $\epsilon$ and $\epsilon',$ of course). Let us now 
consider the intervals $[0,\epsilon]$ and $[\epsilon',\infty[$ separately:
\begin{description}
\item[On {$[\epsilon',\infty[$}:]    
We have    
\[    
0\le m(r) \equiv M    
\,.    
\]    
Then    
\[    
\frac{1}{s}\le \frac{1}{\sqrt{s(s-2m(s))}} \equiv    
\frac{1}{\sqrt{s(s-2M)}} \quad\forall r\in [\epsilon',\infty]    
\,.    
\]    
As in this interval $r \ge R,$ we find, with Eq.~(\ref{def_g}),   
\[    
\ln\left(\frac{r}{R}\right) = \int_{R}^r\frac{ds}{s} 
\le g(r) \equiv     
\int_{R}^r\frac{ds}{\sqrt{s(s-2M)}} =     
2\ln\left(     
\frac{\sqrt{r}+\sqrt{r-2M}}{\sqrt{R}+\sqrt{R-2M}}     
\right)     
\,.    
\]    
Inserting it into Eq.~(\ref{sol_h}), yields (since $\exp$ is a 
monotonically increasing function)
\begin{equation}\label{p_epsilonp_large}    
\frac{h(R)}{R}\, r \le h(r)    
\equiv h(R)\left(\frac{\sqrt{r}+\sqrt{r-2M}}{\sqrt{R}+\sqrt{R-2M}}\right)^2    
\le \frac{4h(R)}{\left(\sqrt{R}+\sqrt{R-2M}\right)^2}\, r    
\,.  
\end{equation}
Note especially that 
$\lim_{r\to\infty}\frac{h(r)}{r}=\frac{4h(R)}{\left(\sqrt{R}+\sqrt{R-2M}\right)^2},$ 
but, by asymptotic flatness, $\lim_{r\to\infty}\frac{h(r)}{r} = 1;$
therefore $h(R)=\frac{1}{4}\left(\sqrt{R}+\sqrt{R-2M}\right)^2>0.$ 
Thus, $h(R)>0,$ and, from Eq.~(\ref{p_epsilonp_large}),
\begin{equation}\label{epsilonp_large}    
0<    
\frac{\left(\sqrt{R}+\sqrt{R-2M}\right)^2}{4h(R)}    
\le \chi(r) \le    \frac{R}{h(R)}    
<\infty    
\quad\forall r\in [\epsilon',\infty[
\,.    
\end{equation}
\item[On {$[0,\epsilon]$}:]    
We have    
\[    
0\le m(r) = 4\pi\int_0^r \varepsilon_0(s) s^2\, ds    
\le \frac{4 \pi}{3}\hat\varepsilon_0 r^3 =:\frac{c_0}{2} r^3    
\,,    
\]    
where $\hat\varepsilon_0 := \sup_{r\in [0,R]}\varepsilon_0(r)> 0.$    
Now, choose $\epsilon>0,$ such that $1-c_0 r^2 > 0$ on $[0,\epsilon]$ \   
(e.g.,~$\epsilon:=(2\sqrt{c_0})^{-1}$).    
Then    
\[    
\frac{1}{s}\le \frac{1}{\sqrt{s(s-2m(s))}} \le    
\frac{1}{s\sqrt{1-c_0 s^2}}    
\quad\forall r\in [0,\epsilon]    
\,,    
\]    
and, since in this interval $r \le R,$ we find, with Eq.~(\ref{def_g}),    
\[    
\ln\left(\frac{R}{r}\right) = \int_r^{R}\frac{ds}{s} \le -g(r) \le     
\int_r^{R}\frac{ds}{s\sqrt{1-c_0 s^2}} = \ln\left(\frac{R}{r}\right)     
+\,\ln\left(\frac{1+\sqrt{1-c_0 r^2}}{1+\sqrt{1-c_0 R^2}}\right)     
\,.    
\]    
Again, inserting it into Eq.~(\ref{sol_h}), yields 
\[    
\frac{h(R)}{R}\, r \ge h(r) \ge  
\frac{h(R)}{R}\frac{1+\sqrt{1-c_0 R^2}}{1+\sqrt{1-c_0 r^2}}\, r    
\ge \frac{h(R)\left(1+\sqrt{1-c_0 R^2}\right)}{2R}\, r    
\,,    
\]    
and, hence,
\begin{equation}\label{epsilon_small}    
0<\frac{R}{h(R)} \le \chi(r) \le    
\frac{2 R}{h(R)\left(1+\sqrt{1-c_0 R^2}\right)}<\infty    
\quad\forall r\in [0,\epsilon]
\,.    
\end{equation}
\end{description}
We can therefore conclude that,
since $\R_0^+ = [0,\epsilon] \cup [\epsilon,\epsilon'] \cup [\epsilon',\infty[$ 
and $\chi$ is bounded (from above and below) in each of these subintervals, 
$\chi$ is bounded (from above and below) in $\R_0^+.$ 
\hbox{} \hfill $\square$

\subsection*{\normalsize \ The function $H$}

\bigskip

\noindent
We have
$$
H(r) :=\frac{-e^{\frac{-\lambda (r)}{2}}\,
          [-6 + 6\,e^{\frac{\lambda (r)}{2}} + r\,\nu '(r)]
          }{2\,h(r)} \,,
$$
and
\begin{eqnarray*}
e^{-\lambda(r)/2}
& = &
\sqrt{1-\frac{2m(r)}{r}}
\\
r \nu'(r)
& = &
\frac{2m(r) + 8\pi r^3 p_0(r)}{r-2m(r)}
=
\frac{\frac{2m(r)}{r} + 8\pi r^2 p_0(r)}{1-\frac{2m(r)}{r}}
\,.
\end{eqnarray*}
Thus,
\begin{equation}\label{fH}
H(r) = \frac{1}{2h(r)}
\Big\{ 
6\left[\sqrt{1-\frac{2m(r)}{r}} - 1\right] \negthickspace
- \left[\frac{2m(r)}{r}+ 8\pi r^2 p_0(r)\right]
\left(\negthickspace \sqrt{1-\frac{2m(r)}{r}} \, \right)^{\negthickspace -1}
\negthickspace \Big\}
\,.
\end{equation}
Now, using the Cauchy-Schwarz inequality in (\ref{fH}),
and the following estimates in $r\in [0,\epsilon],$
for some $\epsilon \in ]0,R[$ small, (see former Subsec.
in Appendix A)
\begin{eqnarray}
&&  0 \le m(r) \le \frac{c_0}{2}r^3 \\
&&  0 \le \sqrt{1 - x}  \le 1 - \frac{x}{2}
\quad\forall x\in [0,1]\\
&& 0\le c_1 r  \le h(r) \le c_2 r \\
&& 0 \le p_0(r) \le \hat{p}_0 := \sup_{r\in[0,R]}p_0(r)
\quad\forall r\ge 0
\,,
\end{eqnarray}
where the constants $c_{i \ (i=0,\ldots,2)}$ are all strictly positive
(and finite), we get
\begin{eqnarray}
|H(r)| & \le &
\frac{1}{2h(r)}
\Big\{ 
6\left|\sqrt{1-\frac{2m(r)}{r}} - 1\right|
+ \left|\frac{2m(r)}{r}+ 8\pi r^2 p_0(r)\right| \negthickspace
\left(\negthickspace \sqrt{1-\frac{2m(r)}{r}}\, \right)^{\negthickspace -1}
\negthickspace \Big\}
\nonumber 
\\
& \le & 
\frac{1}{2c_1 r}
\Big\{ 
6 \left[\frac{c_0}{2}r^2\right]
+ \left[c_0 r^2 + 8\pi\hat{p}_0 r^2\right]
(\sqrt{1-c_0\epsilon^2})^{-1}
\Big\} 
\nonumber
\\
& =: &
\label{estimate_H}
\frac{c_3 r^2}{c_1 r} =: c_4 r
\,,
\end{eqnarray} 
with $0<c_3,\,c_4<\infty.$ 
Therefore $H$ is bounded in $[0,\epsilon].$ (Especially, 
due to Eq.~(\ref{estimate_H}), $H(0)=0.)$
And, since, by Eq.~(\ref{fH}), $H$ is also continuous in the open interval
$]0,\infty[$ and $\lim_{r\to\infty}H(r) = 0$ 
(because $\lim_{r\to\infty} \frac{h(r)}{    r} = 1),$
$H$ is bounded everywhere in $[0,\infty[.$
\hbox{}\hfill$\square$

\setcounter{equation}{0}
\renewcommand{\theequation}{B \arabic{equation}}

\section*{\normalsize APPENDIX B: \ The minimum principle for generalized
supersolutions}

Consider in a domain (open and connected set) $G \subset \R^n$ \ 
$(n \ge 2)$ the differential operator with principal part of 
divergence form, defined by
$$
L u = \partial_{i} [a_{ij}(x) \partial_{j} u + a_i(x) \, u]
+ b_i(x) \, \partial_{i} u + c(x) \, u \, ,
$$
with $a_{ij}=a_{ji}.$ Notice, an operator $L$ of the general form 
$L u = \tilde{a}_{ij}(x) \partial_{ij} u + \tilde{b}_i(x) \partial_i u + \tilde{c}(x) u$ 
may be written in divergence form provided its principal coefficients $\tilde{a}_{ij}$ are
differentiable. If furthermore the $\tilde{a}_{ij}$ are constants, then 
even with coinciding coefficients $(a_{ij} = \tilde{a}_{ij},$ $b_i = \tilde{b}_i,$ 
$c = \tilde{c})$ and $a_i \equiv 0.$ \ Let us assume that
\begin{enumerate}

\item[1.] $L$ is strictly elliptic in $G,$ i.e. 
$\exists$ a constant $\lambda > 0$ such that $\lambda \le$
the minimum eigenvalue of the principal coefficient matrix $[a_{ij}(x)],$ 
\begin{equation}
\label{st_ell}
\lambda \, |y|^2 \le a_{ij}(x) \, y_i y_j \ \ \ \ \forall y \in \R^n,
\ \ \forall x \in G \, ;
\end{equation}
\item[2.] $a_{ij},$ $a_i,$ $b_i,$ and $c$ are measurable
and bounded functions in $G,$ 
\begin{equation}
\label{b}
|a_{ij}| < \infty, \quad |a_i| < \infty, \quad
|b_i| < \infty, \quad |c| < \infty \ \ \mbox{in} \ G \quad
(i,j \in \{1,\ldots,n\}) \, .
\end{equation}
\end{enumerate}

By definition, for a function $u$ which is only assumed to be {\em weakly differentiable}   
and such that the functions \, $a_{ij} \partial_{j} u + a_i u$ \, and \, 
$b_i \partial_{i} u + c u,$ $i=1,\ldots, n$ \ are locally integrable (in particular, for 
$u$ belonging to the Sobolev space $W^{1,2}(G)),$ \ $u$ is said to satisfy
$Lu = g$ in $G$ {\em in a generalized (or weak) sense} 
$(g$ also a locally integrable function in $G)$
if it satisfies
\begin{eqnarray*}
{\cal L} (u, \varphi; G) & := & \int_{G} \{ (a_{ij} \partial_{j} u + a_i u) 
\partial_{i} \varphi - (b_i \partial_{i} u + c u) \varphi \}
d x \\
& = & - \int_{G} g \, \varphi \, d x, \ \ \ \ \
\forall \varphi \ge 0 \ \ \ \varphi \in C^1_c(G) \, 
\end{eqnarray*}
(where $C^1_c(G)$ is the set of functions in $C^1(G)$ with compact support in $G).$

\vspace{1mm}

Notice, $u$ is {\em generalized supersolution} relative to a
differential operator $L$ and the domain $G$ (i.e. satisfies $Lu \le 0$ in $G$ 
in a generalized sense) if it satisfies 
${\cal L} (u, \varphi; G) \ge 0, \ 
\forall \varphi \ge 0 \ \ \varphi \in C^1_c(G).$

\vspace*{4mm}

\noindent {\bf Theorem 1:} (weak minimum principle)\\
{\it Let $u \in W^{1,2}(G),$ \ $G$ a bounded domain, satisfy $Lu \le 0$
in $G$ in a generalized sense with 
\begin{equation}
\int_{G} (c \varphi - a^i \partial_i \varphi) \, d x \le 0, \ \ \  
\forall \varphi \ge 0 \ \ \ \varphi \in C^1_c(G) \, .
\label{c<0}
\end{equation}
\noindent and conditions {\em (\ref{st_ell})} and {\em (\ref{b})} above,\\
then
$$
\min_{\overline{G}} u \ge \min_{\partial G} u^{-}, \ \ \ \ \ (u^{-} \equiv \min(u,0) \, ) \, .
$$
}

\vspace*{2mm}

(A proof of this theorem can be found in Ref.~13, Theorem 8.1.)
\bigskip

\noindent {\bf Theorem 2:} (strong minimum principle)\\
{\it Let $u \in W^{1,2}(G) \cap C^0(G)$ satisfy $Lu \le 0$ in $G$ in a generalized sense,
with the operator $L$ satisfying conditions {\em (\ref{st_ell})}, {\em (\ref{b})}, 
and {\em (\ref{c<0})},\\
then
$u$ cannot achieve a non-positive minimum in the interior of $G,$ unless $u \equiv$ const.
}

\vspace*{2mm}

(A proof of this theorem can be found in Ref.~13, Theorem 8.19.)
Note that the weak minimum principle, Theorem 1, for $C^0(G)$ supersolutions
is a direct consequence.

\bigskip


\end{document}